\documentclass[a4paper,11pt]{article}
\pdfoutput=1 

\usepackage{jcappub} 

\usepackage[T1]{fontenc} 

\usepackage{natbib}
\title{Evaluating the Accuracy of Reionization Prescriptions in Semi-analytic Models of the First Stars and Galaxies}

\author[a,b]{Thomas Behling}
\emailAdd{tbehlin@rockets.utoledo.edu}

\author[a]{, Ryan Hazlett}
\emailAdd{ryan.hazlett@rockets.utoledo.edu}

\author[a,c]{, Mihir Kulkarni}
\emailAdd{mihir.kulkarni@uni-goettingen.de}

\author[a]{, Eli Visbal}
\emailAdd{elijah.visbal@utoledo.edu}

\affiliation[a]{Department of Physics and Astronomy and Ritter Astrophysical Research Center, University of Toledo, 2801 W. Bancroft Street,
Toledo, Ohio 43606, USA}

\affiliation[b]{Department of Physics and Astronomy, Michigan State University, 567 Wilson Rd,  East Lansing, MI 48824, USA}

\affiliation[c]{Institut für Astrophysik und Geophysik, Georg-August Universität Göttingen, Friedrich-Hund-Platz 1, D-37077 Göttingen, Germany}

\abstract{Semi-analytic models are a valuable tool to study the first stars and galaxies. Their numerical efficiency makes it possible to survey broad regions of astrophysical parameter space across large volumes and redshift ranges. Following reionization in these models is necessary since star formation is suppressed in ionized regions due to photoheating of the gas. Here we evaluate the accuracy of three semi-analytic reionization prescriptions (two previously developed and one new model) by comparing their three-dimensional distribution of ionized bubbles to the \emph{Renaissance} hydrodynamical cosmological radiative transfer simulations. 
We find that the previously existing models accurately determine the distribution of the larger bubbles within our ${\sim}6$ comoving Mpc simulation box, but that these models fail to take into account self-shielded neutral gas in dense filaments. Thus, these prescriptions overestimate the fraction of halos in HII regions impacted by reionization feedback by up to an order of magnitude (depending on halo mass and redshift). This leads to an unrealistically large effect of reionization feedback on Pop III stars and low-mass metal-enriched galaxies. Our newly developed model takes into account the density structure of the cosmic web, leading to good agreement with \emph{Renaissance} in the fraction of halos found in ionized regions.
}

\begin{document}
\maketitle
\flushbottom

\section{Introduction} \label{sec:intro}
Understanding the earliest stages of galaxy evolution is currently an exciting frontier in astronomy. There are a variety of observations that can or will soon probe primordial Population III (Pop III) stars (for a review see \cite{2023ARA&A..61...65K}). These include possible direct detections from the James Webb Space Telescope (JWST) \cite{2024A&A...687A..67M, 2025arXiv250111678F, 2023arXiv230514413V}, 21cm cosmology \cite{2018MNRAS.478.5591M, 2022MNRAS.516..841G}, absorption systems found in high-redshift quasar spectra \cite{Sodini2024}, stellar archaeology in the local universe \cite{Sodini2024}, HeII line intensity mapping \cite{2015MNRAS.450.2506V, 2022ApJ...933..141P}, the optical depth of the cosmic microwave background (CMB) \cite{2015MNRAS.453.4456V}, and supernovae or gamma ray bursts from Pop III progenitors \cite{2016SSRv..202..159T, 2018MNRAS.479.2202H, 2019PASJ...71...59M}.

Theoretical models are required to interpret these measurements. There are a variety of different modeling techniques, each with their own strengths. For example, hydrodynamical cosmological simulations \cite[e.g., ][]{1999ApJ...527L...5B, 2015MNRAS.448..568H, 2015ComAC...2....3G} can follow detailed physical processes with high fidelity, but are numerically expensive. Semi-analytic modeling is a complementary approach \cite{2018MNRAS.474..443G, 2020MNRAS.493.1217M, 2020visbal, asloth, 2023MNRAS.525..428H, 2024ApJ...962...62F}. Here dark matter halo merger histories are first generated, typically with either cosmological N-body simulations or Monte Carlo (MC) methods exploiting the extended Press-Schechter (EPS) formalism \cite{1974ApJ...187..425P, 1991ApJ...379..440B, 1993MNRAS.262..627L}. Then analytic prescriptions are used to predict star formation within the dark matter halos. Semi-analytic models have the advantage of being very numerically efficient while still capturing the important physical processes of galaxy formation, and can be calibrated to agree well with hydrodynamical simulations \cite{2018ApJ...859...67C, 2025ApJ...978...13H}. Their efficiency enables semi-analytic models to survey the relevant parameter space (e.g., initial mass function (IMF) of Pop III stars), which is not feasible with hydrodynamical cosmological simulations due to their numerical expense.

Theoretical models must include a number of feedback effects to accurately model Pop III stars and early galaxies. For example, Lyman-Werner (LW) radiation produced by stars and galaxies photodissociates molecular hydrogen, which delays subsequent Pop III star formation (since H$_2$ is the primary coolant in the formation of Pop III stars) \cite{1997ApJ...476..458H, 2001ApJ...548..509M, 2008ApJ...673...14O,2014MNRAS.445..107V, 2021ApJ...917...40K}. It is also necessary to follow the process of cosmic reionization. This is because gas in ionized regions is photoheated to ${\sim}10^4~{\rm K}$, which suppresses star formation is low-mass dark matter halos \cite{1998MNRAS.296...44G, 1996ApJ...465..608T, 2004MNRAS.348..753S, 2004ApJ...601..666D, 2006MNRAS.371..401H, 2014MNRAS.444..503N}. Other key processes include inhomogeneous metal enrichment \citep[e.g., ][]{2015MNRAS.452.2822S} and the impact of the baryon-dark matter streaming velocities \cite{2010PhRvD..82h3520T, 2013MNRAS.432.2909F}.

In this paper, we focus on understanding the accuracy of semi-analytic techniques modeling the early stages of reionization relevant to Pop III stars and high-redshift galaxy formation. Two previous semi-analytic models have implemented prescriptions of inhomogeneous reionization in this context following the three-dimensional (3D) distribution of HII regions and their impact on Pop III stars (A-SLOTH \cite{asloth} and \cite{2020visbal} which we refer to as V2020 throughout the remainder of this paper). As described in detail below, these models use relatively simple prescriptions to generate a distribution of spherical ionized regions around sources of ionizing radiation. We note that these models are used to compute reionization in boxes of length ${\sim}3~{\rm Mpc}$. This is too small of a region to accurately model the global process of reionization \cite{2014MNRAS.439..725I}, but it is necessary to accurately compute Pop III and early galaxy formation \cite{2020visbal}.

We test the accuracy of these models by comparing them directly to the \emph{Renaissance Simulations} \cite{2015ApJ...807L..12O, 2016ApJ...833...84X}, which are hydrodynamical cosmological simulations including star formation (Pop III and metal-enriched), supernovae feedback, and radiative transfer. 
This is related to a previous study comparing semi-analytic models of reionization to radiative transfer simulations \cite{2011MNRAS.414..727Z}, but that test was performed on much larger scales and comparing different methods. We note that these authors refer to  the approximate reionization schemes as ``semi-numerical'', but we use the term semi-analytic throughout this paper. We find that when assuming the same sources of ionizing photons as \emph{Renaissance}, the V2020 and A-SLOTH models recreate a similar large-scale distribution of ionized bubbles, but fail to capture self-shielding within neutral dense filaments in the cosmic web. This motivates us to create a new semi-analytic model of reionization that includes the 3D density field (using the only the dark matter component from the simulation). This leads for much better performance than the previous models.

The rest of this paper is structured as follows. In Section \ref{sec:renaissance}, we describe the \emph{Renaissance Simulations}. In Section \ref{sec:methods}, we discuss the previous semi-analytic reionization schemes from V2020 and A-SLOTH. We also introduce our new method. In Section \ref{sec:results}, we present a direct comparison between these semi-analytic methods and \emph{Renaissance}. Finally, we summarize our results and conclude in Section \ref{sec:conclusions}. Throughout we assume a cosmology consistent with \emph{Renaissance} using parameters from the WMAP7 $\Lambda$CDM+SZ+LENS best fit \citep{2011ApJS..192...18K} with $\Omega_{\mathrm{M}} = 0.266$, $\Omega_{\mathrm{\Lambda}} = 0.734$, $\Omega_{\mathrm{b}} = 0.0449$, $h = 0.71$, $\sigma_{\mathrm{8}} = 0.81$, and $n = 0.963$.

\section{Renaissance Simulations} \label{sec:renaissance}
The \emph{Renaissance Simulations} are a set of hydrodynamical cosmological simulations focused on the first stars and galaxies  \cite{2015ApJ...807L..12O, 2016ApJ...833...84X}. They were run with the adaptive mesh refinement code (AMR) {\sc enzo} \cite{2014ApJS..211...19B}, and include Pop III and metal-enriched star formation, supernovae feedback (including the pollution of halos and the intergalactic medium (IGM) with metals via supernovae winds), and radiative feedback. Radiative transfer is performed with a ray-tracing code, {\sc moray} \cite{2011MNRAS.414.3458W}. Primordial chemistry and cooling is computed for nine species of H and He \cite{1997NewA....2..181A} and metallicity dependent cooling is also included \cite{2009ApJ...691..441S}. In this paper, we utilize the ``Normal'' simulation, which is a zoomed region with a volume of ($6 \times 6\times 6.125~{\rm Mpc}^3$) within a larger $(40~{\rm Mpc})^3$ box selected to be near the mean density of the universe. The dark matter mass resolution is $2.9\times10^4~M_\odot$ and the maximum level of refinement corresponds to a maximum resolution of 19 comoving pc for the gas.
We use the LW background presented in \cite{2016ApJ...833...84X} that self-consistently calculates a LW background from star formation in the Renaissance Normal region.
As described in more detail below, when comparing the semi-analytic models to \emph{Renaissance}, we take the stars found in \emph{Renaissance} 
at the available simulation snapshots ($z=15.7$, $14.7$, $13.7$, $12.7$, and $11.6$) and use them to compute the distribution of ionized bubbles predicted by the semi-analytic models. This distribution is then compared to the neutral fraction in the radiative transfer simulation.

\section{Semi-analytic Reionization Schemes} \label{sec:methods}
In this section, we describe three semi-analytic models of reionization. The first two were previously implemented in V2020 \cite{2020visbal} and A-SLOTH \cite{asloth}, and the third is newly developed here.
Each of these models takes in the sources of ionizing photons and then determines the 3D ionization state throughout a simulation box.  After introducing the models, we discuss the details of how we compare them to the radiative transfer simulations of \emph{Renaissance}.

\subsection{V2020} \label{subsec:visbal2020}
The V2020 model \cite{2020visbal} is a modified version of the excursion-set approach utilized in previous work to model reionization on much larger spatial scales \cite{2004ApJ...613....1F, 2007ApJ...669..663M,2011MNRAS.411..955M}. In summary, this model places spheres with a range of different radii throughout the box and those which contain more ionizing photons than hydrogen atoms (computed assuming the mean cosmic density) are set to be HII regions.
Operationally, this is performed on a 3D uniform spatial grid. First, the number density of hydrogen-ionizing photons that escape into the IGM from each cell is computed on the grid. This is given by

\begin{equation}
\label{eqn:photon_count}
P_{\rm cell} = \frac{f_{\rm esc,II}\eta_{\rm II} M_{\rm II,cell}}{m_{\rm p}V_{\rm cell}} + 
           \frac{f_{\rm esc,III}\eta_{\rm III}M_{\rm III,cell}}{m_{\rm p}V_{\rm cell}},
\end{equation}
where $f_{\rm{esc,II}}$ is the escape fraction of ionizing photons in dark matter halos hosting metal-enriched Population II (Pop II) stars, $\eta_{\rm{II}}$ is the number of ionizing photons produced per baryon of Pop II stars, $M_{\rm II,cell}$ is the mass of Pop II stars within that cell, $m_{\rm p}$ is the mass of a proton, and $V_{\rm cell}$ is the comoving volume of each cell (the same quantities with subscript III refer to those for Pop III stars).  
This grid of photon density is then smoothed by convolving with a spherical top-hat window function for a variety of bubble scales. Following each smoothing iteration, cells are identified where the number density of smoothed photons exceeds the mean number density of neutral hydrogen. These identified cells are treated as the centers of spherical ionized regions.

An additional correction factor of 1/4 is applied to the photon counts, $P_{\rm cell}$, calculated above. This is due to geometrical effects. Without a correction, this scheme overestimates the size of a bubble around a single source by a factor of 8. However, for multiple sources within a bubble, this overestimate is less. A 1/4 correction results in matching the total number of predicted ionizations and the total number of ionizing photons from sources to within ${\sim}10\%$. 
We applied a set of 30 logarithmically spaced bubble radii, ranging from a single cell to the size of the box, when implementing the model in this paper. We note that hydrogen recombination is ignored in this scheme and the two others described below. Recombination is found to have a relatively minor impact at the redshifts explored here because the rate of new ionizing photon sources is so high. By the time a region could recombine, a much larger number of ionizing photons were produced in the same region.

\subsection{A-SLOTH} \label{subsec:asloth}
A-SLOTH \cite{asloth} predicts the 3D ionization field by first assigning each star-forming halo a spherical ionized bubble, calculated from the size of the region it can ionize given the number of ionizing photons escaping into the IGM (assuming the mean cosmic hydrogen number density throughout the simulation box). When the center of one bubble lies within the ionized radius of another bubble, the two are combined into a new bubble. The new bubble is located at the center of mass of the two parent bubbles and ionizes a region equal to the sum of the parent bubbles' volumes.    

\subsection{New Model} \label{subsec:new_model}
As we show in the following section, the reionization prescriptions from both V2020 and A-SLOTH reasonably reproduce the large-scale bubble distributions of \emph{Renaissance}, but they fail to account for dense filamentary regions that run through the center of some bubbles. In these filaments, the gas often remains neutral. This is particularly important to capture since it is where many dark matter halos are found and one must correctly identify if they are in ionized photoheated regions where star formation is suppressed.
To address this issue, we have developed a new model that leverages the density distribution extracted from N-body simulations (rather than just the halo positions as in V2020 and A-SLOTH).
    
In the V2020 model, the centers of ionized bubbles are obtained for each bubble radii by finding all grid cells where the following condition is true
\begin{equation}
\mathcal{F}^{-1}\{\mathcal{F}\{P(\vec{r})/4\}\mathcal{F}\{k(a)\}\} > \bar{n}_H.
\end{equation}
Here $\mathcal{F}$ denotes a Fourier transform (performed as Fast Fourier Transforms (FFTs) in our implementation), $P(\vec{r})/4$ is the corrected ionization density on the grid, $\vec{r}$ is the position of a particular grid cell, $k(a)=\left(\frac{4}{3}\pi a^3\right)^{-1}$ is a normalized spherical top-hat window function with bubble radius $a$, and $\bar{n}_H$ is the average number density of neutral hydrogen. Note that this is simply using the convolution theorem to smooth the photon density cube. The ionization distribution in the simulation grid, $I(\vec{r})$ , is then determined by placing bubbles around their centers.

In contrast, for our newly constructed model, the ionization distribution is determined by setting $I(\vec{r})=1.0$ where the following relation is satisfied (and $I(\vec{r})=0$ elsewhere):
\begin{equation}
\mathcal{F}^{-1}\{\mathcal{F}\{P(\vec{r})\}\mathcal{F}\{k\}\} > \bar{n}_H(\delta(\vec{r})+1).
\end{equation} 
Here $\delta({\vec{r}})\equiv\frac{\rho(\vec{r})-\bar{\rho}}{\bar{\rho}} $ is the matter overdensity on the grid determined from the N-body simulation being used by the semi-analytic model. The kernel, $k$, is no longer a spherical top-hat, but is instead given by $k(r) = \left(\frac{4}{3}\pi r^3\right)^{-1}$.
In summary, this model operates by first smoothing the photon number density using a new kernel and then comparing the density of photons at each cell against the density of neutral hydrogen (using the simulation's matter density), and setting the ionization fraction to one where the smoothed photon density exceeds the hydrogen density. We chose the kernel such that at each radius it has the value of a spherical top-hat whose edge lies at that radius. This is then similar to applying spherical window functions at many different scales simultaneously. 

A final operation is applied, where the regions of low density with $\delta(\vec{r})<-0.85$ are set to neutral. This represents approximately 1\% cells within the simulation space and overwhelmingly consists of cells that are highly isolated from star-forming halos. Due to the smoothing operations applied, these cells would become ionized despite their isolation. Applying this final operation modestly improves the visual appearance of the model, but has no significant effect on other metrics explored below. Thus, it could be removed in the event that it interferes with model operation at lower redshifts than explored in this work.

To better understand the behavior of our new model, we examined a toy example of a photon source in a simple axisymmetric density field. The overdensity is given by $\delta = 15 / (r^3)$, where $r$ is the distance from a central axis (which runs along the x direction and falls within a $256^3$ cubic grid covering a $(6.14~{\rm Mpc)^3}$ volume region). The overdensity is bounded to a maximum value of $\delta = 15$. This density distribution roughly resembles a straight filament of gas. 
The photon source is assumed to be in the center of the filament and to have emitted $3.65 \times 10^{65}$ hydrogen-ionizing photons (recombinations are ignored).

\begin{figure*}[t]
    \centering
    \includegraphics[width=1\linewidth]{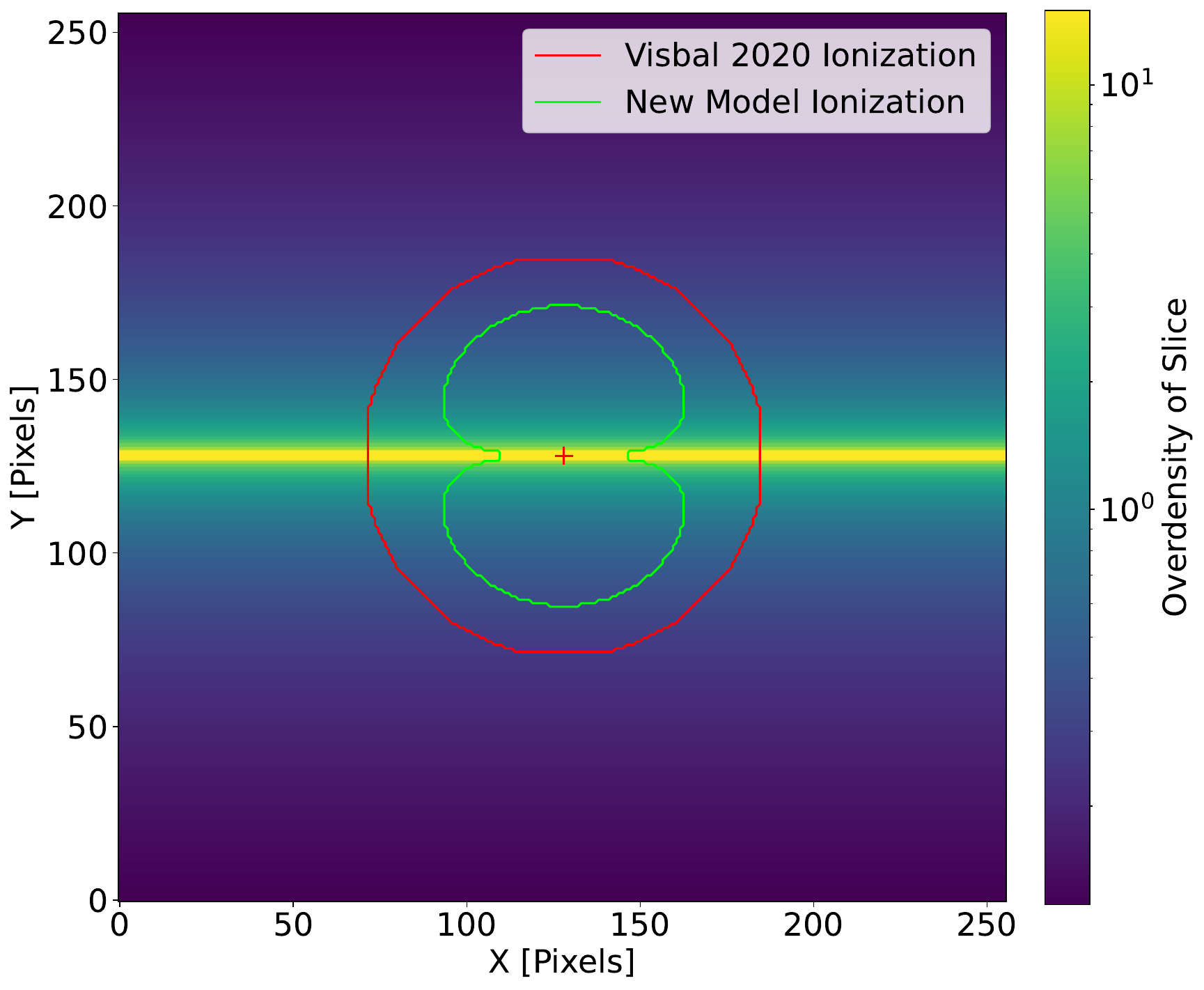}
    \caption{An axisymmetic toy example illustrating the behavior of the V2020 reionization prescription and our new model. We plot a slice centered on an artificial filament described by $\delta = 15 / (r^3)$, where $r$ represents the distance in grid cells from the center of the filament, bounded to a maximum of $\delta = 15$. The location of the ionization source is denoted by the red ``$+$'' and has emitted $3.65\times10^{65}$ ionizing photons. The region is a cube $6.14$ comoving Mpc across.
    The contours show the boundary of the ionized regions produced by each model. As described in the main text, the new model performs better in the sense that it ionizes preferentially into the lower density regions perpendicular to the filament.}
    \label{fig:toy_examples}
\end{figure*}

In Figure \ref{fig:toy_examples}, we compare the ionization surrounding this toy filament for both the V2020 reionization scheme and the new model. We find that the new scheme creates an ionized region that escapes preferentially perpendicular to the filament, as opposed to an unphysical spherical bubble produced by V2020.
We note that in both cases the number of ionized hydrogen atoms implied by the HII regions is much larger than the number of ionizing photons created by the source (a factor of ${\sim}200$ higher in V2020 and a factor of ${\sim}100$ higher in the new model). In V2020, this error is due to the assumption that the gas being ionized is at the mean density of the universe (i.e., the density of the filament is ignored). In the new model, the density of the filament is taken into account leading to a more realistic shape of the HII region, but the new model still ignores important radiative transfer effects such as shadowing and self-shielding. Despite these shortcomings, we find that when more realistic cosmological distributions of the density field are assumed, excellent agreement is obtained (as presented in the following section). This is in part due to the fact that we treat the Pop III and Pop II escape fractions as free parameters, which reduces the strength of the ionizing sources. We also note that ionization and supernovae feedback reduce the density near sources, which is not captured in this toy example. 

\subsection{Renaissance Comparisons} \label{subsec:impl_details}
Before presenting how the three semi-analytical models compare to \emph{Renaissance} in the next section, we will now describe the details of how this comparison was implemented.
Each model was run using a spatial resolution of $256^3$, in a box $6.14$ comoving Mpc across (this was chosen to completely encompass the \emph{Normal} region in \emph{Renaissance}. We assume that the Pop II and Pop III stellar masses that enter the model are those taken directly from \emph{Renaissance} data (described in Section \ref{sec:renaissance}). We also assume that the numbers of ionizing photons produced per baryon in Pop III and Pop II stars in our semi-analytic runs are given by $\eta_{\rm III}=65000$ and $\eta_{\rm II}=4000$. Note that the number of ionizing photons in our A-SLOTH and new model is computed following Eq.~\ref{eqn:photon_count}.

We calibrate each semi-analytic model separately to \emph{Renaissance} by treating the escape fractions $f_{\rm esc,III}$ and $f_{\rm esc,II}$, as free parameters.
For each model, a ratio between the global ionization fractions from the model and \emph{Renaissance} was obtained at each of the snapshot redshifts and their absolute deviations from 1 were summed together. We define this sum as our loss function, and the resulting fits are determined by minimizing it for each semi-analytic model.
When performing this fit, we only use data from the $128^3$ cell region in the center of the simulation.
We do not include the outskirts of the box because there are a number of strong ionizing sources that created bubbles extending well beyond our simulation data cube complicating the relation between stellar mass and ionized fraction found within the cube.
Finally, we note that in V2020 the N-body simulations used had periodic boundary conditions, while the \emph{Normal} region in \emph{Renaissance} does not.  In order to apply our convolutions without periodic boundary conditions for V2020 and the new model here, we pad our simulation regions within empty boxes of double their original size before convolving with FFTs.

\section{Results} \label{sec:results}
\begin{figure*}[t]
    \centering
    \includegraphics[width=1\linewidth]{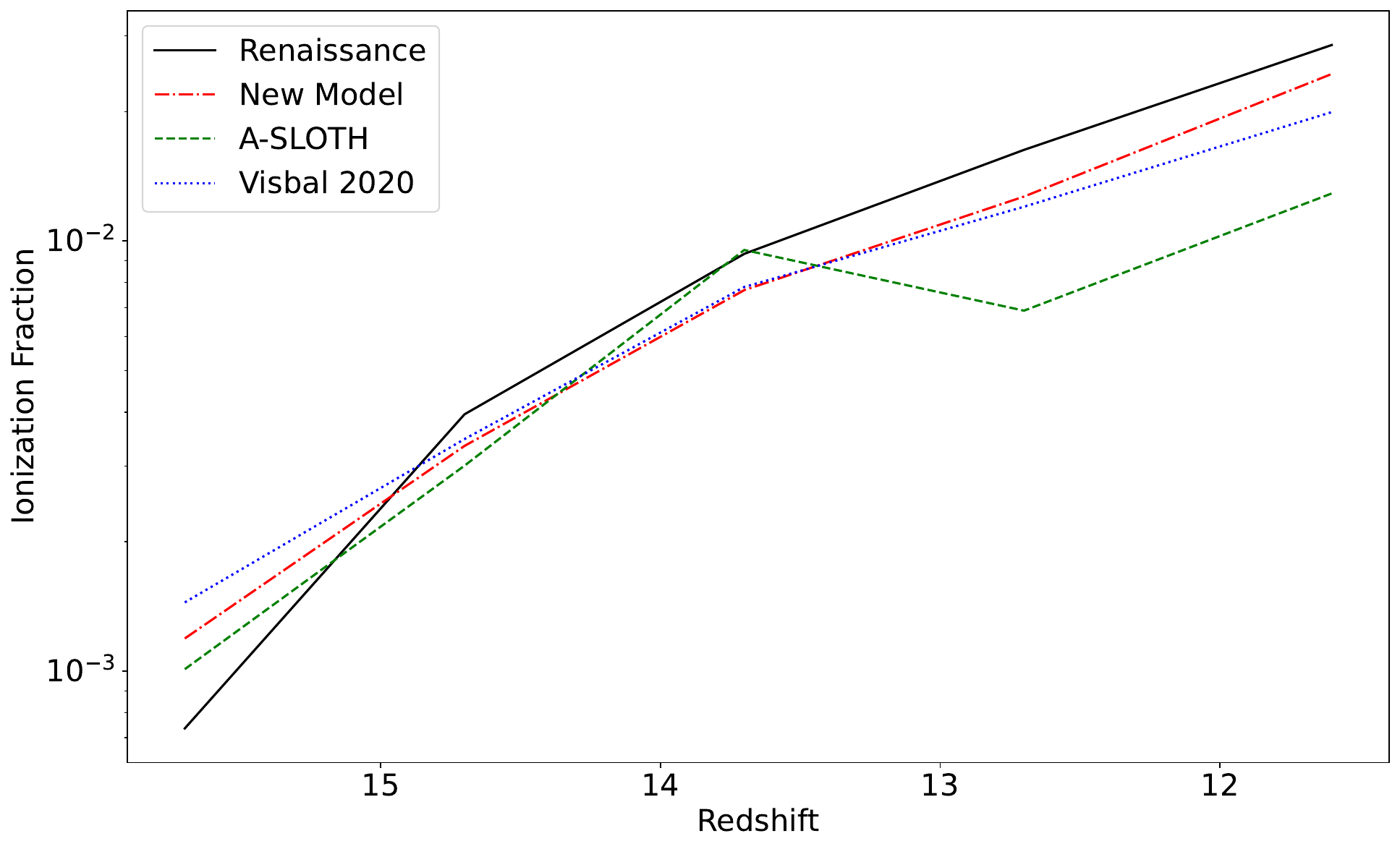}
    
    \caption{The total simulation box ionization fractions from the best fit Pop II and Pop III escape fractions in our semi-analytic models and for \emph{Renaissance}. }
    
    \label{fig:escape_fractions}
\end{figure*}

In this section, we evaluate the performance of each of the three semi-analytic reionization models described above by comparing them with the \emph{Renaissance} radiative transfer simulation. First, we calibrated the models by finding the best fit values of $f_{\rm esc, II}$ and $f_{\rm esc, III}$ using the method described in \ref{subsec:impl_details}. Figure \ref{fig:escape_fractions} shows the global ionization fraction in each calibrated model compared to the values from \emph{Renaissance}. The best-fit escape fractions (which are assumed to be constant with redshift) and the global ionization fractions in our final snapshot redshift are shown in Table \ref{tab:escape_fractions}. We find agreement in the ionized fraction to within a factor of a few across all redshifts, but note that we do not expect perfect agreement. In the simulation, there is substantial variation from halo to halo in the escape fraction and a more complicated relationship between escape fractions and halo masses than assumed in the semi-analytic models \cite{2016ApJ...833...84X}. We note that our best-fit escape fractions in all three semi-analytic models are roughly a factor of a few less than the mean escape fractions reported in \emph{Renaissance} \cite{2016ApJ...833...84X}. This is consistent with the results of the toy model presented in the previous section. The semi-analytic models tend to create HII regions that are too large, which is compensated for by this lower escape fraction in the calibration.

\begin{table*}[t]
    \begin{tabular}{|c|c|c|c|}
        \hline
         Model & $f_{\rm esc, II}$ &  $f_{\rm esc, III}$ & Box Ionization Fraction ($z = 11.6$) \\
        \hline\hline
         \emph{Renaissance} Selection & - & - & $2.855\%$ \\
         \hline
         Visbal 2020 & $1.538\%$ & $16.0\%$ & $1.993\%$ \\
         \hline
         A-SLOTH & $1.00\%$ & $21.0\%$ & $1.290\%$ \\
         \hline
         New Model & $1.75\%$ & $11.0\%$ & $2.445\%$ \\
         \hline
    \end{tabular}
    \centering
    \caption{The calibrated values of the Pop II and Pop III escape fractions for each model and the overall ionization fractions of each simulation at the final simulation snapshot.}
    \label{tab:escape_fractions}
\end{table*}

\begin{figure*}[t]
    \includegraphics[width=1\linewidth]{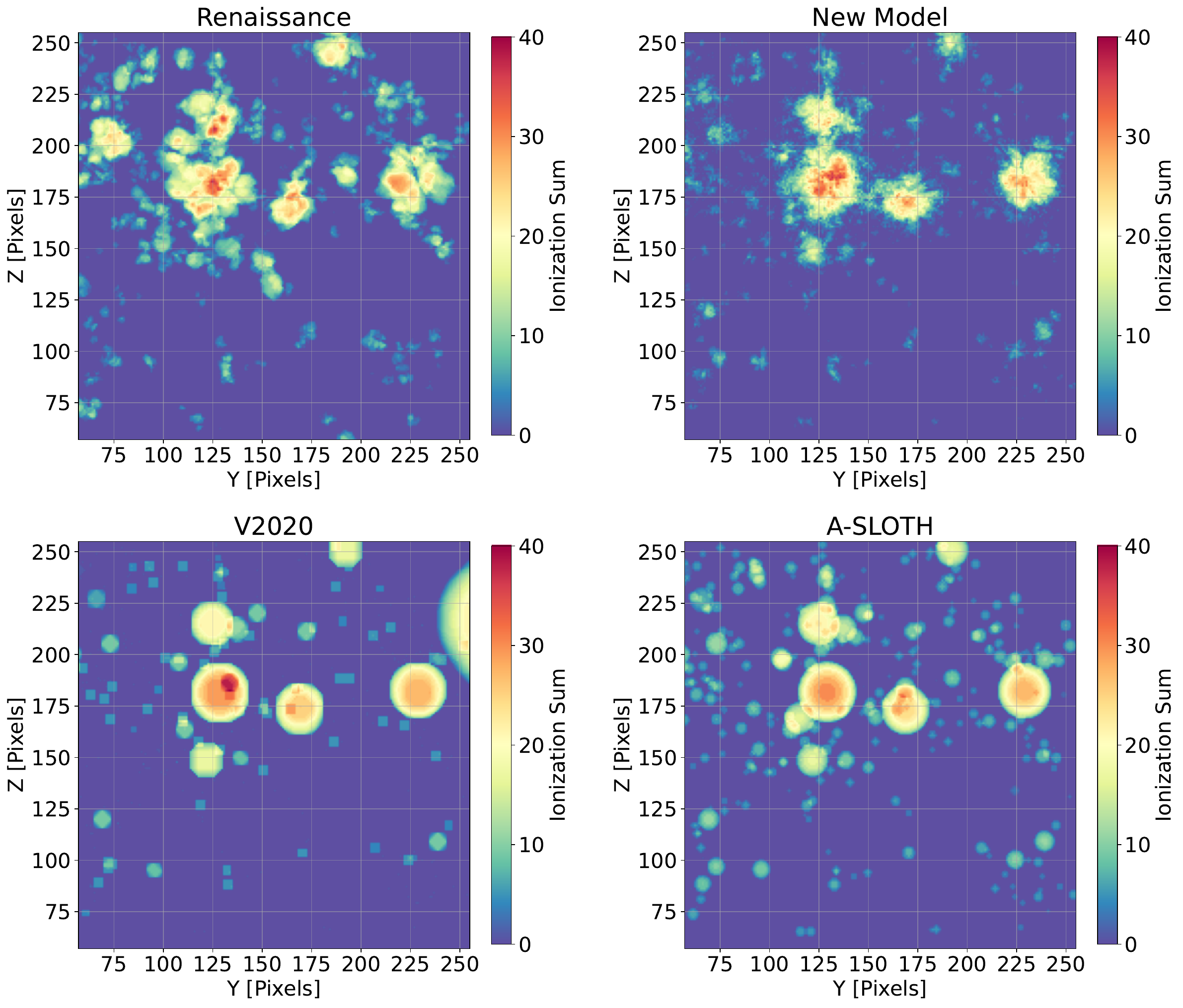}

\caption{Projections of the ionization distribution through a large subset of our simulation (corresponding to ${\sim}6$ Mpc on a side). These projections are taken along the x-axis at a depth of 150 cells. The colors show the sum of the ionized cells along the x-axis.}

\label{fig:fullbox_slices}
\end{figure*}

\begin{figure*}[t]
\includegraphics[width=1\linewidth]{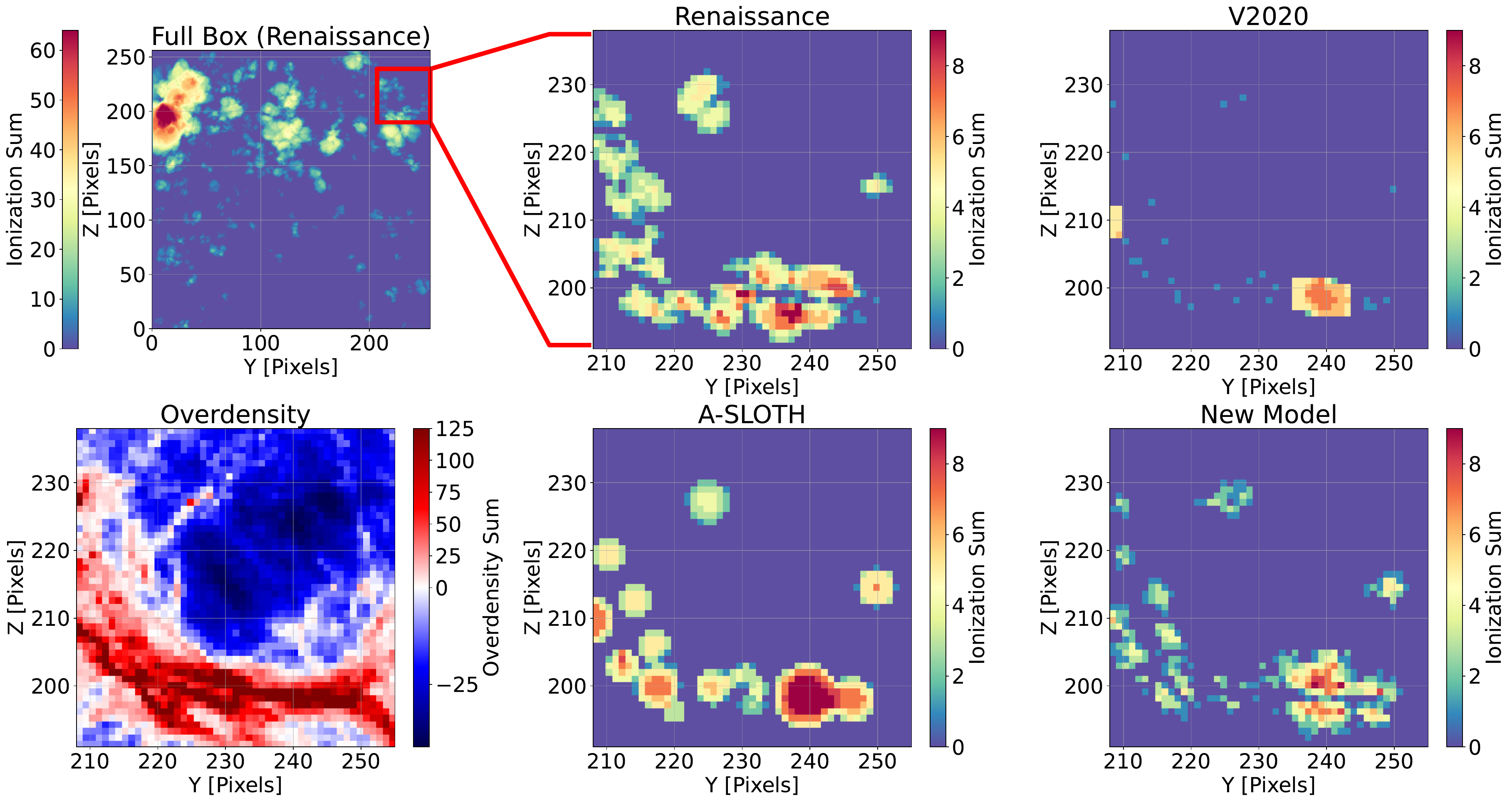}

\caption{Projections through a subsection of the ionization distribution in each of the 4 referenced models, as well as the overdensity ($\delta (\vec{r})$) distribution extracted from Renaissance. These projections are taken along the x-axis of a 51 cell cube (corresponding to ${\sim}$1 Mpc) from within the larger simulation space. Additionally, a projection with a depth of 150 cells through Renaissance is included, with the location of the chosen subset highlighted (upper leftmost panel). Note that overdensity sum and ionization sum refer to the projected sums of $\delta$ and the ionization fraction along the x-axis.}

\label{fig:representative_slices}
\end{figure*}

In Figure \ref{fig:fullbox_slices}, we show projections of the ionization field through a large-scale portion of the simulation box for both the semi-analytic models and \emph{Renaissance}. For the largest HII regions, all models agree relatively well with the simulations, but there are significant differences on smaller scales. A-SLOTH and V2020 assume perfectly smooth spherical bubbles (though some appear cube-like in V2020 due to finite grid resolution), missing the finer structure seen near the edges of ionized regions in the radiative transfer simulation. Because our new method takes into account small-scale fluctuations in the matter density field, it is able to better mimic this behavior.

We can see this more clearly in Figure \ref{fig:representative_slices}, where we show the projection of a zoomed-in region roughly 1 comoving Mpc in scale containing an overdense filament in the cosmic web. Stars and ionized bubbles form preferentially in and around this filament in \emph{Renaissance}. However, because the center of the filament is very dense ($\delta{\sim}100$ in dark matter), the gas is able to self-shield and/or recombine and remains neutral. This is completely missed in V2020 and A-SLOTH, where spherical bubbles ionize these dense regions. On the other hand, our new model captures the neutral strip along the center of the filament. As we discuss next, this has important implications for feedback on star formation due to reionization. We note that the precise locations of the ionized regions in the new model are not identical to those in \emph{Renaissance}. This is expected since the escape fraction varies from halo to halo in the simulation but not in the semi-analytic model, causing the sources of ionization to be similar but not precisely the same.

We emphasize that perhaps the most important reason to follow reionization in semi-analytic models of the first stars and galaxies is to determine where star formation is prevented by photoheating of the gas \cite{1998MNRAS.296...44G, 1996ApJ...465..608T, 2004MNRAS.348..753S, 2004ApJ...601..666D, 2006MNRAS.371..401H, 2014MNRAS.444..503N}. For example, V2020 assumes that within HII regions star formation is delayed or strongly suppressed in halos with a virial mass below $1.5\times10^8 \left ( \frac{1+z}{11} \right )^{-1.5}$. We investigated how well the semi-analytic models identify halos impacted by this effect by examining the fraction of halos without stars that are in ionized regions and comparing them to \emph{Renaissance}. We consider halos without stars since star-forming halos rapidly self-ionize their surroundings and automatically agree in all models and the simulation. Star-forming halos also represent a very small fraction of the total number of halos. 

\begin{figure*}[t]
\centering 
\includegraphics[width=12cm]{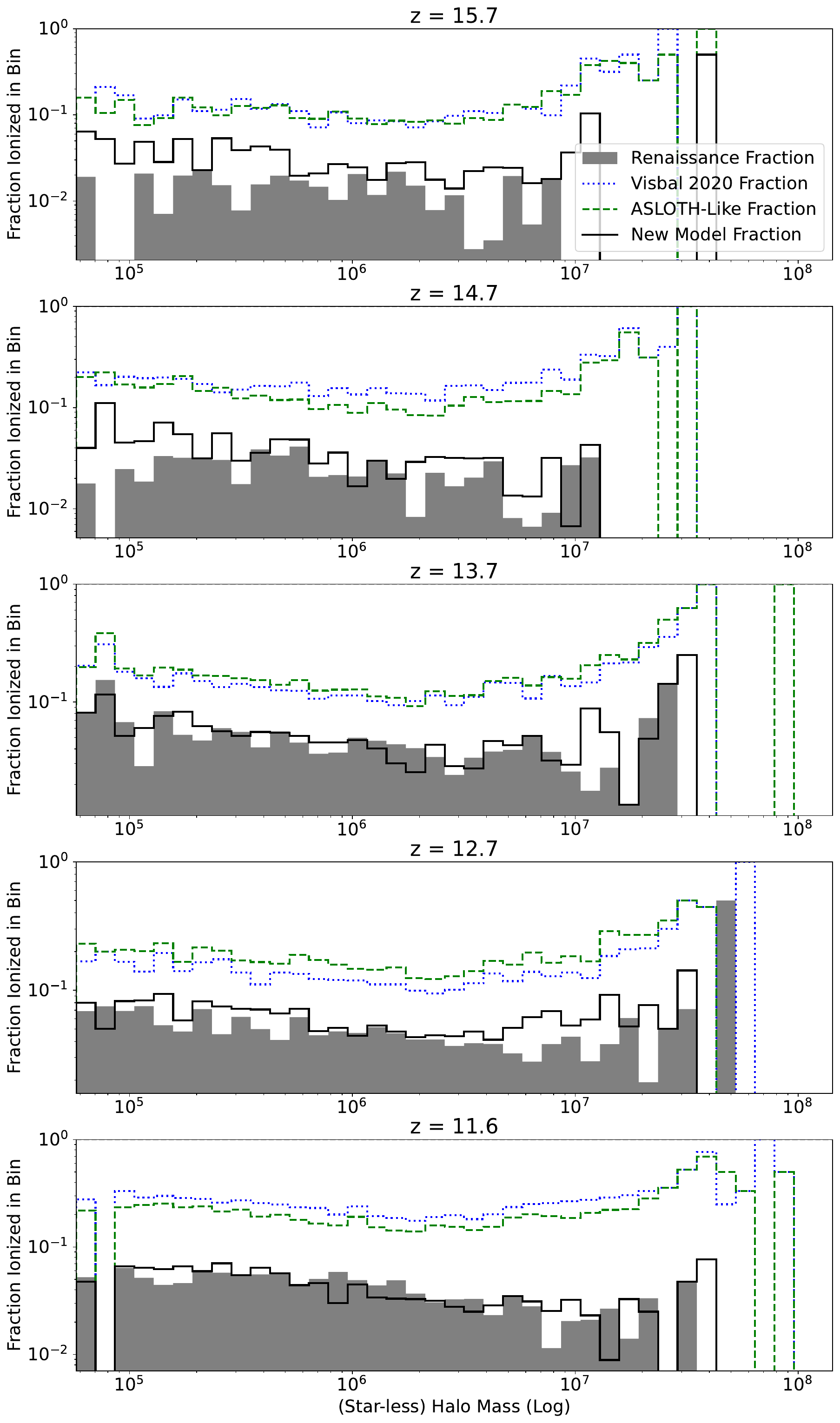}
\caption{The fraction of starless halos within each mass bin that are ionized for each of the models. We see that previous semi-analytic models overestimate the fraction of ionizied halos, whereas the new model much more closely mimics the radiative transfer simulations of \emph{Renaissance}. To give a sense for the number of halos, we note that in the $z=11.6$ panel we find that the bin at $10^5 M_\odot$ contains $406$ halos, the bin at $10^6 M_\odot$ contains $1383$ halos, and the bin at $10^7 M_\odot$ contains $476$ halos.}
\label{fig:starless_ionization}
\end{figure*}

We present the results of this analysis in Figure \ref{fig:starless_ionization}, which shows the fraction of starless halos in ionized regions as a function of halo mass. We find that both the V2020 and A-SLOTH reionization models lead to substantially more halos in ionized regions compared to \emph{Renaissance} (between a factor of a few and an order of magnitude depending on halo mass and redshift). This is because their simple spherical bubbles improperly ionize dense self-shielded regions within the cosmic web as shown in Figure \ref{fig:representative_slices}. Our new method has up to an order of magnitude improvement in the agreement with the ionized fraction. It shows excellent agreement with \emph{Renaissance} except in some of the highest halo mass bins where there are very few halos. This is due to the new model's ability to take into account small-scale density fluctuations. Our results suggest that previous methods are significantly overestimating the importance of ionization feedback on the first stars, since many halos are embedded in dense neutral pockets and avoid photoheating.
    
\section{Summary and Conclusions}
\label{sec:conclusions}
In this paper, we evaluated the performance of three prescriptions for reionization in small-scale semi-analytic models of the first stars and galaxies (i.e., with box sizes of a few comoving Mpc). The 3D ionization fields predicted by these models were compared directly to the \emph{Renaissance} cosmological hydrodynamical radiative transfer simulations. The stars in each model were taken from the simulation and the escape fractions of ionizing radiation from halos into the IGM were calibrated to match the global ionization fraction from \emph{Renaissance}. Two of the models explored, V2020 \cite{2020visbal} and A-SLOTH \cite{asloth}, were previously developed for studies that focused on Pop III star formation. V2020 utilizes an excursion-set approach assuming a uniform IGM, while A-SLOTH applies spherical HII regions around sources and combines overlapping bubbles around their center of mass. We also tested a new scheme developed here, which similarly to V2020 is built on a 3D grid. The new prescription first smooths a grid with the number density of ionized photons by convolving with a kernel $k(r) = \left(\frac{4}{3}\pi r^3\right)^{-1}$. The smoothed grid is then compared with the inhomogeneous hydrogen density field (estimated from the dark matter distribution in the simulation) and regions where the ionizing photon density exceeds the hydrogen number density are assumed to lie within HII regions. 

We find that all three semi-analytic models successfully reproduce the approximate large-scale 3D distribution of HII regions. However, the V2020 and A-SLOTH models perform poorly on smaller scales. In \emph{Renaissance}, dense gas along filaments in the cosmic web remains neutral due to self-shielding. Generally speaking, the spherical bubbles of V2020 and A-SLOTH erroneously ionize these dense regions. Our new model however, is able to capture this effect more closely matching the radiative transfer simulation. 

This has important implications for reionization feedback due to photoheating of gas which suppresses star formation in low-mass halos. We find that both V2020 and A-SLOTH overpredict the number of starless halos found within ionized regions (by up to an order of magnitude depending on halo mass and redshift, see Figure \ref{fig:starless_ionization}). On the other hand, our new model agrees very well with the number of ionized halos in \emph{Renaissance}. This is because halos are able to remain self-shielded within dense gas that is not taken into account in the older methods. This suggests that previous methods will underpredict the amount of star formation in low mass halos, both for Pop III and Pop II stars.

Our findings motivate several new lines of research. First, we have only tested our new method to a redshift of $z=11.6$ corresponding to a global ionized fraction in \emph{Renaissance} of ${\sim}0.03$. In future work, we intend to compare our new algorithm to the results of the \emph{THESAN} \cite{2022MNRAS.511.4005K} simulations to determine its accuracy in larger simulations and at lower redshifts. We also note that we did not perform an exhaustive study on the impact of different choices for the functional form of the kernel. 
Finally, we intend to run this new scheme within updated versions of V2020 to see how it changes the Pop III and Pop II star formation history when accurate small-scale reionization and star formation are simultaneously followed self-consistently.  

\begin{acknowledgments}
This work was supported by NSF grant AST-2009309 and NASA ATP grant 80NSSC22K0629. The computations were performed at the Ohio Supercomputer Center. We thank John Wise for his assistance with the \emph{Renaissance Simulations}.
\end{acknowledgments}

\bibliography{main}

\end{document}